\documentclass[pra,aps,reprint,showpacs,superscriptaddress,amsmath]{revtex4-1}

\usepackage{bm}
\usepackage{graphicx}
\usepackage{dsfont}

\newcommand{\bra}[1]{\langle #1 | \,}
\newcommand{\ket}[1]{\, | #1 \rangle}

\newcommand{\expv}[1]{\langle #1 \rangle}
\newcommand{\om}{\omega}
\newcommand{\Om}{\Omega}
\newcommand{\ga}{\gamma}
\newcommand{\Ga}{\Gamma}
\newcommand{\de}{\delta}
\newcommand{\De}{\Delta}
\newcommand{\ka}{\kappa}
\newcommand{\la}{\lambda}
\newcommand{\La}{\Lambda}

\newcommand{\bx}{\mathbf{x}}
\newcommand{\db}{d_{\mathrm{b}}}
\newcommand{\dcb}{d_{\mathrm{cb}}}
\newcommand{\dr}{d_{\mathrm{r}}}
\newcommand{\hlf}{\frac{1}{2}}
\newcommand{\mc}[1]{\mathcal{#1}}
\newcommand{\sig}{\hat{\sigma}}
\newcommand{\Sig}{\hat{\Sigma}}
\newcommand{\sigav}{\bar{\sigma}}
\newcommand{\hrho}{\hat{\rho}}
\newcommand{\hS}{\hat{S}}

\begin{document}

\title{Spatial correlations of Rydberg excitations 
in optically driven atomic ensembles}
 
\author{David Petrosyan}
%\email{david.petrosyan@iesl.forth.gr}
\affiliation{Institute of Electronic Structure and Laser, 
FORTH, GR-71110 Heraklion, Crete, Greece}

\author{Michael H\"oning}
\affiliation{Fachbereich Physik und Forschungszentrum OPTIMAS,
Technische Universit\"at Kaiserslautern, D-67663 Kaiserslautern, Germany}

\author{Michael Fleischhauer}
\affiliation{Fachbereich Physik und Forschungszentrum OPTIMAS,
Technische Universit\"at Kaiserslautern, D-67663 Kaiserslautern, Germany}

\date{\today}

\begin{abstract}
We study the emergence of many-body correlations in the stationary state 
of continuously-driven, strongly-interacting dissipative system. 
Specifically, we examine resonant optical excitations of Rydberg states of 
atoms interacting via long-range dipole-dipole and van der Waals potentials
employing exact numerical solutions of the density matrix equations 
and Monte-Carlo simulations.  
Collection of atoms within a blockade distance form a ``superatom'' 
that can accommodate at most one Rydberg excitation. The superatom 
excitation probability saturates to $\frac{1}{2}$ for coherently driven 
atoms, but is significantly higher for incoherent driving, approaching 
unity as the number of atoms increases. In the steady state of 
uniformly-driven, extended one-dimensional system, the saturation 
of superatoms leads to quasi-crystallization of Rydberg excitations 
whose correlations exhibit damped spatial oscillations.
The behavior of the system under the van der Waals interaction potential
can be approximated by an analytically soluble model based on a ``hard-rod''
interatomic potential. 
\end{abstract}


\maketitle

\section{Introduction}

Strong, long-range dipole--dipole (DD) or van der Waals (vdW) interactions 
between atoms in highly excited Rydberg states \cite{RydAtoms} can suppress 
multiple Rydberg excitations within a certain interaction (blockade) 
volume, while enhancing the rate of single collective excitation 
\cite{Lukin2001,Tong2004,Singer2004,Vogt2006,Heidemann2007,UrbanGaetan2009,Dudin2012NatPh}.
The resulting dipole blockade constitutes the basis for a number of promising 
quantum information schemes \cite{rydrev} and interesting many-body effects 
involving long-range correlations and crystallization of Rydberg excitations 
\cite{Weimer2008,Low2009,Weimer10,Pupillo2010,Schachenmayer2010,Sela2011,Pohl2010,Schwarzkopf2011,Schauss2012,Bijnen2011,Viteau2011,Lesanovsky2011,Ji2011,Lesanovsky2012,Ates2012,Loew2012,Garttner2012,Hoening2012}.

Much of the research on strongly interacting Rydberg atoms is focused on 
the unitary dynamics or eigenstates of the many-body system. In coherently 
driven ensembles of atoms, depending on the strength and detuning of driving 
lasers, different ground-state phases with crystalline order emerge 
\cite{Weimer2008,Low2009,Weimer10,Schachenmayer2010,Lesanovsky2011,Sela2011}. 
Yet, adiabatically attaining the ground state of a large system requires 
exceeding preparation times \cite{Pohl2010,Bijnen2011} during which the 
decoherence and dissipation associated with the optical excitation cannot 
be neglected. This necessitates the consideration of open systems and their 
stationary states \cite{Hoening2012}, which is the main purpose of this paper. 

We study theoretically resonant optical excitations of Rydberg states
of atoms interacting with each other via DD and vdW potentials. 
Atoms within a blockade distance form a ``superatom'' which can 
accommodate at most one collective Rydberg excitation 
\cite{Lukin2001,Dudin2012NatPh,Robicheaux2005,Stanojevic2009,Honer2011}. 
We show that the steady-state excitation probability of the superatom 
saturates to $\frac{1}{2}$ for coherently driven atoms, but can be 
significantly higher in the presence of strong dephasing which suppresses 
inter-atomic coherences and disentangles the atoms \cite{Honer2011}. 
This case is amenable to an iterative Monte-Carlo sampling algorithm
which can accurately and efficiently simulate the stationary state 
of the many-body system. We apply this algorithm to extended one-dimensional
(1D) atomic ensembles, and explore both the low-density (lattice)
regime \cite{Schauss2012,Hoening2012} with one or few atoms per blockade 
distance, and the high-density (continuous) regime \cite{Schwarzkopf2011}
with many atoms per blockade distance. 
For the vdW interacting atoms at high densities, tight packing of 
superatoms leads to quasi-crystallization of Rydberg excitations, 
similar to the predictions of an analytic model involving a ``hard-rod'' (HR) 
interatomic potential \cite{Ates2012}. In contrast, the DD potential
appears to be too ``soft'' for assigning a well-defined interaction range 
at high atomic densities.

The paper is organized as follows.
In Sec.~\ref{sec:mathform} we introduce the Hamiltonian and dissipative terms
of the master equation for the density matrix of the many-body system and 
examine the properties of superatoms under coherent and incoherent driving. 
In Sec.~\ref{sec:1dsys} we study extended 1D systems, employing
exact solution of the density matrix equations for several atoms, 
and Monte Carlo simulations for hundreds of atoms.
A model based on the HR interatomic potential is analytically solved in
Sec.~\ref{sec:hr} and compared with the results of numerical simulations
for the vdW interacting atoms. Experimental considerations and 
conclusions are summarized in Sec.~\ref{sec:conclud}

\section{The many-body system}
\label{sec:mathform}

\begin{figure}[t]
\includegraphics[width=7.0cm]{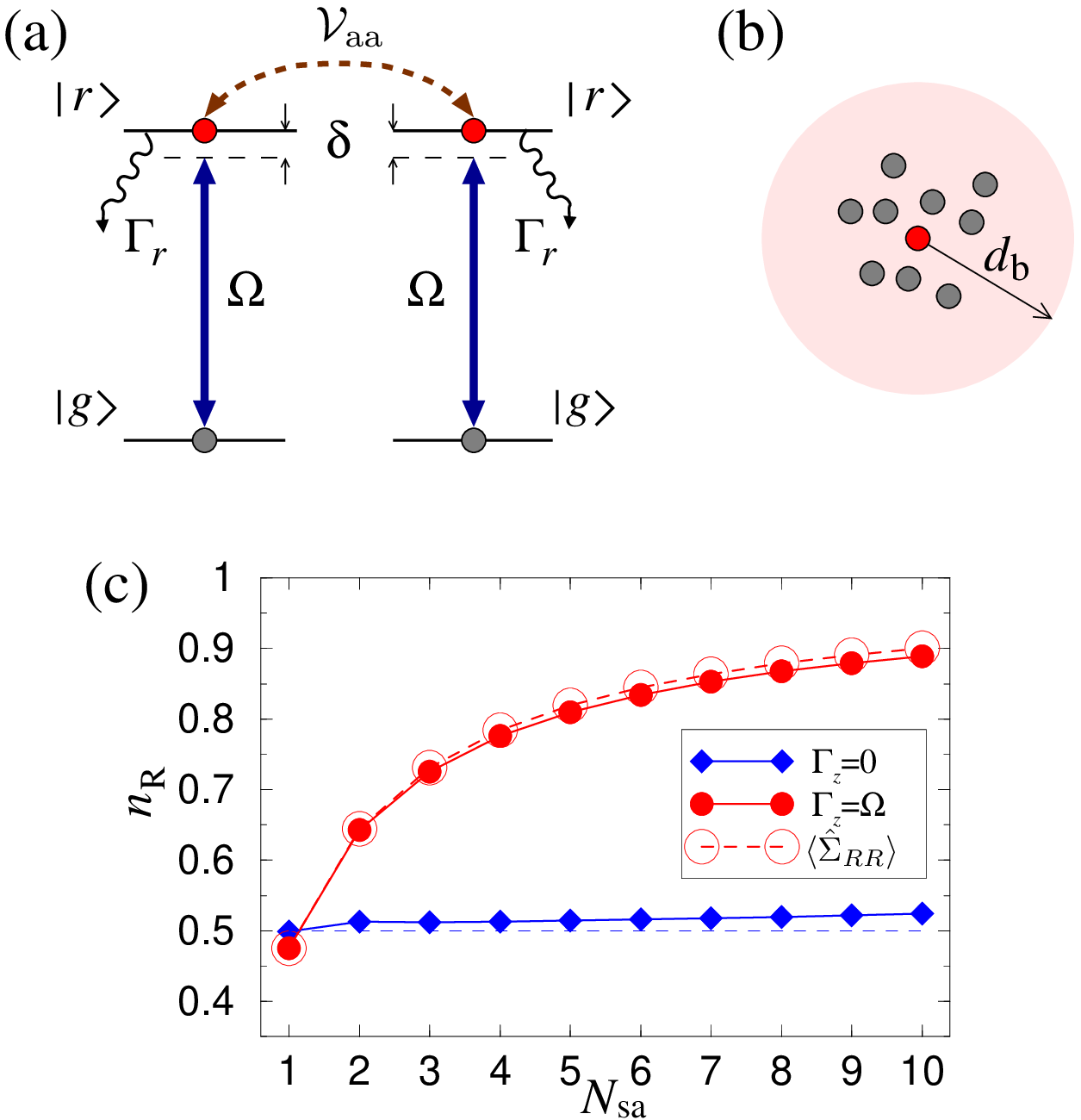}
\caption{%(Color online)
(a)~Level scheme of atoms interacting with the driving field $\Om$ 
on transition $\ket{g} \to \ket{r}$ with detuning $\de$, while 
$\mc{V}_{\mathrm{aa}}$ denotes the DD or vdW interaction between the atoms 
in Rydberg state $\ket{r}$ having the (population) decay rate $\Ga_r$.
(b)~Atoms within the blockade distance $\db$ form a superatom 
with at most one Rydberg excitation.
(c)~Mean number of Rydberg excitations $n_{\mathrm{R}}$ of superatom 
containing $N_{\mathrm{sa}} \leq 10$ atoms within $L =0.7\db$ (vdW interaction), 
obtained from the exact steady-state solutions of Eq.~(\ref{rhoME}) 
for $\de =0$, $\Ga_r = 0.1 \Om$, and $\Ga_z = 0$ (filled blue diamonds) 
and $\Ga_z = \Om$ (filled red circles).
Also shown the Rydberg excitation probability $\expv{\Sig_{RR}}$ 
of superatom (empty red circles) as per Eq.~(\ref{SigRR}). } 
\label{fig:alsSA}
\end{figure}

We consider an ensemble of $N$ atoms irradiated by a uniform driving 
field that couples near-resonantly the atomic ground state $\ket{g}$ to 
the highly excited Rydberg state $\ket{r}$ with Rabi frequency $\Om$, 
Fig.~\ref{fig:alsSA}(a).   
A pair of atoms $i$ and $j$ at positions $\bx_i$ and $\bx_j$ 
excited to states $\ket{r}$ interact either 
via the DD ($p=3$) \cite{ddpot} or vdW ($p=6$) \cite{rydcalc} 
potential $\hbar \Delta(\bx_i -\bx_j) = \hbar C_{p} |\bx_i -\bx_j|^{-p}$.
In the frame rotating with the driving field frequency $\om$, 
the system Hamiltonian $\mc{H} = \mc{V}_{\mathrm{af}} + \mc{V}_{\mathrm{aa}}$ 
is composed of the atom-field and atom-atom interactions, 
$\mc{V}_{\mathrm{af}} =  - \hbar \sum_{j}^N [\de \sig_{rr}^j + 
\Om (\sig_{rg}^j + \sig_{gr}^j)]$ and 
$\mc{V}_{\mathrm{aa}} = \hbar \sum_{i<j}^N \sig_{rr}^i \Delta(\bx_i -\bx_j) 
\sig_{rr}^j$, where $\sig_{\mu \nu}^j \equiv \ket{\mu}_{jj}\bra{\nu}$ 
is the transition ($\mu \neq \nu$) or projection ($\mu = \nu$) operator 
for atom $j$ at position $\bx_j$, and $\de = \om - \om_{rg}$ is the driving
field detuning. The relaxation processes affecting the atoms include 
the spontaneous (radiative) decay of the excited state $\ket{r}$ with 
rate $\Ga_r$, and the (non-radiative) dephasing of atomic coherence $\sig_{rg}$
with rate $\Ga_z$; the decay rate of the Rydberg state is typically 
small compared to $\Om$, while the physical origins of dephasing include 
non-radiative collisions, Doppler shifts or the excitation laser linewidth.
The decay and dephasing Liouvillians, acting independently on each atom,
are given, respectively, by 
$\mc{L}_r^j \hrho = \hlf \Ga_{r} [2 \sig_{gr}^j \hrho \sig_{rg}^j 
- \sig_{gg}^j \hrho - \hrho \sig_{gg}^j]$ and
$\mc{L}_z^j \hrho = \Ga_{z} [(\sig_{rr}^j - \sig_{gg}^j) \hrho (\sig_{rr}^j 
- \sig_{gg}^j) - \hrho]$ \cite{PLDP2007}. 
The density matrix $\hrho$ of the $N$-atom system obeys the master equation 
\begin{equation}
\partial_t \hrho = -\frac{i}{\hbar} [\mc{H}, \hrho] +  \mc{L} \hrho , 
\label{rhoME}
\end{equation}
with $\mc{L} \hrho =  \sum_j^N (\mc{L}_r^j \hrho + \mc{L}_z^j \hrho)$. 

%%%%%%%%%%%%%%%%FIGURE%%%%%%%%%%%%%%%%
%%%%%%%%%%%%%%%%%%%%%%%%%%%%%%%%%%%%%%%

\subsection{Blockade distance}

For a single two-level atom, the steady-state population of the 
excited state $\ket{r}$ is a Lorentzian function of detuning $\de$,
\begin{equation}
\expv{\sig_{rr}} = \frac{\Om^2}
{2 \Om^2 + \frac{\Ga_r}{2\ga_{rg}} (\ga_{rg}^2 + \de^2) } ,
\label{sigrr}  
\end{equation}
with the width $w = \ga_{rg} \sqrt{4 \Om^2 /\Ga_{r}\ga_{rg} + 1}$,
where $\ga_{rg} \equiv \hlf \Ga_{r} + 2 \Ga_{z}$ is the total
(transversal) relaxation rate of the $\sig_{rg}$ coherence \cite{PLDP2007}. 
For strong ($\Om^2 > \Ga_r \ga_{rg}$) resonant ($\de \ll w$) driving,
the population saturates to $\expv{\sig_{rr}} \to \hlf$. 
But given an atom in the Rydberg state $\ket{r}$, it will induce 
a level shift $\De$---equivalent to detuning $\de$---of another 
atom, blocking its Rydberg excitation when $\De \gtrsim w$. 
We may therefore define the blockade distance $\db$ via $\De(\db) = w$, 
which yields 
\begin{equation}
\db \equiv \sqrt[p]{\frac{C_p}{w}}  
\simeq \left( \frac{C_p}{2 \Om} \sqrt{\frac{\Ga_r}{\ga_{rg}}} \right)^{1/p},
\label{blockd}
\end{equation}
with $p=3$ for the DD interaction and $p=6$ for the vdW interaction.

\subsection{The superatom}

Consider $N_{\mathrm{sa}}$ atoms within distance $L < \db$, such that 
$\De(\bx_i -\bx_j) \gg w$ for any pair of atoms $i$ and $j$, 
Fig.~\ref{fig:alsSA}(b). We thus expect that the ``superatom'' can accommodate
at most one Rydberg excitation \cite{Robicheaux2005,Stanojevic2009}. 
This is confirmed by our exact numerical simulations for 
$N_{\mathrm{sa}} \leq 10$; starting with all the atoms in the ground 
state $\ket{g}$, we propagate Eq.~(\ref{rhoME}) for time $t$ ($\gg \Om^{-1}$) 
long enough until the steady-state is reached. 
In Fig.~\ref{fig:alsSA}(c) we show the resulting mean number 
of Rydberg excitations 
$n_{\mathrm{R}} = \expv{\sum_j^{N_{\mathrm{sa}}} \sig_{rr}^j}$ within 
the superatom. Clearly, $n_{\mathrm{R}} < 1 \, \forall \, N_{\mathrm{sa}}$, 
while we verify that the probabilities of double 
$\expv{\sig_{rr}^i\sig_{rr}^j}$, 
triple $\expv{\sig_{rr}^i\sig_{rr}^j\sig_{rr}^k}$, etc. 
excitations are always small. 

In the absence of dephasing, $\Ga_z = 0$ and 
$\ga_{rg} = \hlf \Ga_{r} \ll \Om$, we have 
$n_{\mathrm{R}} \simeq \hlf$ independent of $N_{\mathrm{sa}}$. 
The ground state $\ket{G} = \ket{g_1,g_2, \ldots,g_{N_{\mathrm{sa}}}}$ of the
superatom is coupled only to the single collective Rydberg excitation 
state $\ket{R^{(1)}} = \frac{1}{\sqrt{N_{\mathrm{sa}}}}
\sum_j^{N_{\mathrm{sa}}} \ket{g_1,g_2, \ldots,r_j,\ldots,g_{N_{\mathrm{sa}}}}$,
while all the states $\ket{R^{(n)}}$ with higher number $n > 1$ 
of Rydberg excitations are shifted out of resonance by the strong 
interatomic interaction $\mc{V}_{\mathrm{aa}}$ and therefore 
are not populated. As the collective Rabi frequency $\sqrt{N_{\mathrm{sa}}}\Om$
saturates the transition $\ket{G} \leftrightarrow \ket{R^{(1)}}$, 
the superatom ground and excited states acquire populations 
$\expv{\Sig_{GG}} \simeq \expv{\Sig_{RR}} \simeq \hlf$, 
where $\Sig_{GG} \equiv \prod_{j=1}^{N_{\mathrm{sa}}} \sig_{gg}^j$ 
and $\Sig_{RR} \equiv \sum_{j=1}^{N_{\mathrm{sa}}} \sig_{rr}^j 
\prod_{i \neq j}^{N_{\mathrm{sa}}} \sig_{gg}^i$ are projectors 
onto the ground and single Rydberg excitation states of 
$N_{\mathrm{sa}}$ atoms. 
We note that $n_{\mathrm{R}}$ slightly larger than $\hlf$ seen in 
Fig.~\ref{fig:alsSA}(c) is due to imperfect blockade of Rydberg excitation 
[$\De(L) \sim 10 w$] of atoms at the boundaries of the region of finite 
size [$L = 0.7\db$].

Remarkably, strong dephasing $\Ga_z$ increases the mean number 
of Rydberg excitations $n_{\mathrm{R}} > \hlf$ within the superatom. 
The transversal relaxation $\ga_{rg} \gtrsim \Om$ destroys the 
inter- and intra-atomic coherences, causing individual atoms to behave 
independently. In the basis of collective, singly-excited states, 
this corresponds to a coupling of the symmetric state $\ket{R^{(1)}}$
to all the non-symmetric states \cite{Honer2011}. 
The superatom still contains at most one Rydberg excitation, 
$\Sig_{GG} + \Sig_{RR} = \mathds{1}$,
which, upon combining with $\sig_{gg}^j + \sig_{rr}^j = \mathds{1}$, 
yields 
\begin{equation}
\expv{\Sig_{RR}} = \frac{N_{\mathrm{sa}} \expv{\sig_{rr}}}
{(N_{\mathrm{sa}} - 1) \expv{\sig_{rr}} + 1 } , \label{SigRR}
\end{equation}
where $\expv{\sig_{rr}}$ is given by Eq.~(\ref{sigrr}).
For $N_{\mathrm{sa}} = 1$ we have $\expv{\Sig_{RR}} = \expv{\sig_{rr}}$ 
as it should, while $N_{\mathrm{sa}} \gg 1/\expv{\sig_{rr}}$ leads to 
$\expv{\Sig_{RR}} \to 1$. In Fig.~\ref{fig:alsSA}(c) we plot 
$\expv{\Sig_{RR}}$, which reproduces well the exact numerical solution, 
$n_{\mathrm{R}} \simeq \expv{\Sig_{RR}}$, under the same conditions.

\section{Extended 1D system}
\label{sec:1dsys}

We next consider the steady-state distribution of Rydberg excitations 
in a 1D system of $N = \rho_{\mathrm{at}} L$ atoms of linear density 
$\rho_{\mathrm{at}}$.

\subsection{Small system}

%%%%%%%%%%%%%%%%FIGURE%%%%%%%%%%%%%%%%
\begin{figure}[t]
\includegraphics[width=8.0cm]{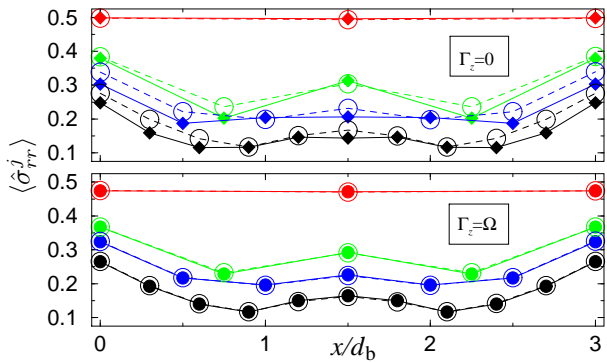}
\caption{%(Color online) 
Rydberg state populations $\expv{\sig_{rr}^j}$ of $N =3,5,7,11$ atoms 
(from top to bottom: red, green, blue, black)
in a 1D lattice of length $L=3 \db$ (vdW interaction) obtained from 
the exact solutions of Eq.~(\ref{rhoME}) for $\de =0$, $\Ga_r = 0.1 \Om$, 
and $\Ga_z = 0$ (upper panel) and $\Ga_z = \Om$ (lower panel).
Also shown the corresponding Rydberg excitation probabilities
$\sigav_{rr}^j$ (empty circles) obtained from MC simulations. } 
\label{fig:1DExSol}
\end{figure}
%%%%%%%%%%%%%%%%%%%%%%%%%%%%%%%%%%%%%%%

We have performed exact numerical simulations
of the density matrix Eqs.~(\ref{rhoME}) for $N \leq 11$ atoms 
with and without dephasing $\Ga_z$. 
In Fig.~\ref{fig:1DExSol} we show the stationary populations 
$\expv{\sig_{rr}^j}$ of Rydberg states of vdW interacting atoms 
in a lattice of length $L = 3 \db$ [open boundary conditions (OBC)]. 
When the interatomic distance exceeds $\db$ ($N=3$), 
interactions play no role and the population $\expv{\sig_{rr}^j}$ 
for each atom is given by Eq.~(\ref{sigrr}). With increasing atomic 
density, interactions progressively suppress the Rydberg
state populations of individual atoms. Simultaneously, we observe an onset
of spatial oscillations of Rydberg excitations $\expv{\sig_{rr}^j}$. 
Note that at higher atomic densities ($N \geq 7$), the oscillations 
are somewhat smoother and more pronounced in the presence of strong 
dephasing $\ga_{rg} \gtrsim \Om$.

\subsection{Monte-Carlo algorithm}

The dephasing suppresses interatomic coherences and disentangles 
the atoms, admitting only classical $N$-body correlations. 
Each atom then behaves as a driven two-level system of Eq.~(\ref{sigrr}) 
but with the detuning $\de$ determined by operator 
$\hS_j \equiv \sum_{i \neq j}^N \sig_{rr}^i \Delta(\bx_i -\bx_j)$ 
which describes the total interaction-induced shift of level $\ket{r}$
for an atom  at position $\bx_j$ involving the contributions of all 
the Rydberg atoms $\sig_{rr}^i$ at positions $\bx_i$. 

We can now introduce an efficient procedure to simulate the stationary 
distribution of Rydberg excitation probabilities at any atomic density 
$\rho_{\mathrm{at}}$. Our algorithm relies on iterative Monte-Carlo (MC) 
sampling of $\{ \sig_{rr}^j \}$ for an ensemble of $N$ atoms, 
in the spirit of the Hartree-Fock method. 
We start with, e.g., all the atoms in the ground state, 
$\expv{\sig_{gg}^j} = 1 \, \forall \, j \in [1,N]$, although the resulting 
steady-state does not depend on the initial configuration.
At every step, for each atom $j$, we draw a uniform random 
number $s \in [0,1]$ and compare it with the Rydberg state 
population $\expv{\sig_{rr}^j}$; if $s \leq \expv{\sig_{rr}^j}$, 
we set $\sig_{rr}^j \to 1$, otherwise $\sig_{rr}^j \to 0$. 
In turn, the thus constructed binary configuration of Rydberg excitations 
$\{ \sig_{rr}^i \} \to \{ 0,1,0,0 \ldots \}$ determines the level shift 
$\hS_j$ (equivalent to detuning $\de$) of atom $j$ when evaluating 
$\expv{\sig_{rr}^j}$. We continuously iterate this procedure, sifting 
repeatedly through every atom in the potential generated by all the other 
atoms, in the self consistent way. The probability distribution 
$\sigav_{rr}^j$ of Rydberg excitations results from averaging over many 
($\sim 10^6$) configurations $\{ \sig_{rr}^j \}$. 
We note a related algorithm employed in \cite{Heeg2012}.
In Fig.~\ref{fig:1DExSol} we plot $\sigav_{rr}^j$ for various $N$ and 
verify that, for strong dephasing ($\Ga_z = \Om$), the algorithm accurately 
reproduces the exact steady-state populations $\expv{\sig_{rr}^j}$. 
In the absence of dephasing ($\Ga_z = 0$), however, the steady-state 
resulting from the MC simulations quantitatively differs from the exact 
state of the system.

\subsection{Large system}

%%%%%%%%%%%%%%%%FIGURE%%%%%%%%%%%%%%%%
\begin{figure}[t]
\includegraphics[width=8.7cm]{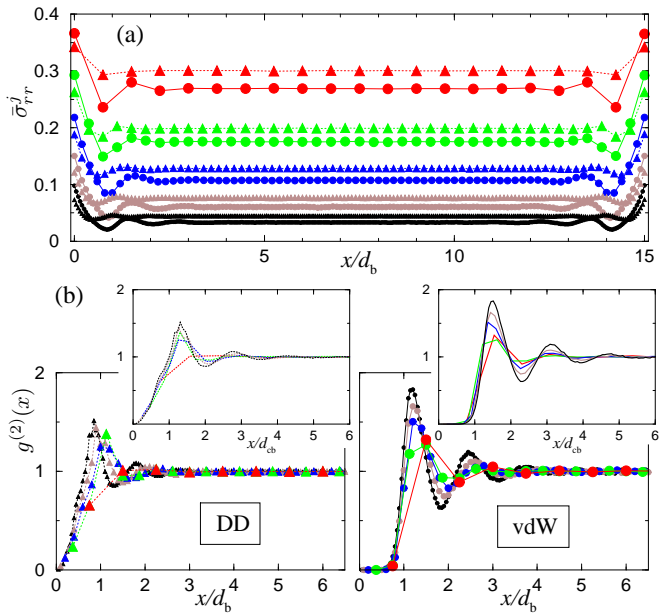}
\caption{%(Color online)
(a) Probabilities $\sigav_{rr}^j$ of Rydberg excitations in a 1D atomic 
ensemble of length $L = 15 \db$ (OBC) obtained from MC simulations 
with $\de =0$, $\Ga_r = 0.1 \Om$, and $\Ga_z = \Om$. 
Triangles (connected by dotted lines) correspond to the DD 
interaction, and circles (connected by solid lines) to the vdW 
interaction between the atoms. The atomic densities are, from top to bottom:
$\rho_{\mathrm{at}} \db = 1\frac{1}{3},2 \frac{2}{3},5,10,20$
(red, green, blue, brown, black).
(b) The corresponding 1D spatial correlations $g^{(2)}(x)$ of 
Rydberg excitations of DD (left) and vdW (right) interacting atoms. 
In the inset of each graph, $g^{(2)}(x)$ is plotted 
versus $x$ in units of the corresponding collective blockade distance 
$\dcb$ of Eq.~(\ref{collblockd}).} 
\label{fig:lattg2}
\end{figure}
%%%%%%%%%%%%%%%%%%%%%%%%%%%%%%%%%%%%%%%

We employ the above MC procedure to study realistically large 1D systems 
of up to $N \sim 10^3$ atoms interacting via the DD and vdW  
interactions and subject to strong dephasing. 
In Fig.~\ref{fig:lattg2}(a) we show the spatial 
distribution of Rydberg excitation probabilities in a finite system 
of length $L = 15 \db$ (OBC) for different atomic densities. 
Interactions between the atoms suppress the Rydberg excitation 
probabilities $\sigav_{rr}^j$, the more the higher is the atomic density 
(smaller the interatomic distance) and stronger the interaction. 
The Rydberg excitations are then repelled to the boundaries of the system. 
An atom at the edge, having higher probability to be excited, suppresses 
the excitation of the neighboring atoms within the blockade distance, 
beyond which another atom acquires higher excitation probability. 
Hence a Rydberg excitation density wave develops. This behavior 
is significantly more pronounced for the vdW interaction between the 
atoms, which is stronger within the blockade distance but falls-off 
fast outside of it, as compared to the DD interaction, which is weaker 
at short distances and has longer tails. The vdW potential is thus more 
reminiscent of a ``hard-rod'' interaction, while the DD potential 
is much ``softer''.   

From many configurations $\{ \sig_{rr}^j \}$, we can extract
the spatial correlations of Rydberg excitations
$G^{(2)}(x \equiv |\bx_i - \bx_j|) 
= \overline{\sig_{rr}^i \sig_{rr}^j}$.  
More explicitly, we place a Rydberg atom ($\sig_{rr}^i = 1$) 
at the origin $\bx_i = 0$, and then use the above MC procedure 
to calculate $\sigav_{rr}^j \, \forall \, \bx_j > 0$, from which we obtain 
the normalized correlation function as 
$g^{(2)}(x) = \sigav_{rr}^j/\sigav_{rr}$,
where $\sigav_{rr} = \frac{1}{N} \sum_{j=1}^N \sigav_{rr}^j$ is the spatial 
average with $N \gg 1$ ($L \gg \db$).
In Fig.~\ref{fig:lattg2}(b) we show the corresponding 
$g^{(2)}(x)$ for the DD and vdW interactions.  
In the case of vdW interaction, the Rydberg atom at $x=0$  
almost completely blocks the excitation of all the atoms within 
the blockade distance $x \lesssim \db$. In contrast, for the DD potential 
the excitation blockage is only partial as $g^{(2)}(x)$
grows nearly linearly in the region of $x < \db$. 
In addition, for higher atomic densities $\rho_{\mathrm{at}} \db > 5$, 
we observe damped spatial oscillations of $g^{(2)}(x)$ with 
increasing amplitude and slowly decreasing period $\la$ close to $\db$.

\subsection{Collective blockade distance}

In a dense system, $\rho_{\mathrm{at}} \db \gg 1$, an atom in the Rydberg state 
$\ket{r}$ blocks the excitation of other atoms within a certain distance 
which, due to collective effects, is somewhat smaller than the blockade 
distance $\db$ for a pair of atoms, Eq.~(\ref{blockd}). 
To estimate the collective blockade distance $\dcb$, note that the 
collective Rabi frequency $\Om_{\mathrm{cb}} = \sqrt{N_{\mathrm{cb}}} \Om$, 
and thereby the excitation linewidth 
$w_{\mathrm{cb}} \simeq 2 \Om_{\mathrm{cb}} \sqrt{\ga_{rg}/\Ga_r}$
for $N_{\mathrm{cb}} = \rho_{\mathrm{at}} \dcb$ atoms is enhanced by 
a factor of $\sqrt{N_{\mathrm{cb}}}$. Substituting $N_{\mathrm{cb}}$
into the definition of 
$\dcb \equiv \sqrt[p]{C_p/w_{\mathrm{cb}}}$ yields
\begin{equation}
\dcb = \left( \frac{C_p}{ \sqrt{\rho_{\mathrm{at}}} w} \right)^{2/(2p+1)}
= \frac{\db}{ (\rho_{\mathrm{at}} \db)^{1/(2p+1)}} , \label{collblockd}
\end{equation}
with $p=3$ for DD and $p=6$ for vdW interactions. In turn, the number
of atoms within the collective blockade distance is 
$N_{\mathrm{cb}} = (\rho_{\mathrm{at}}\db)^{2p/(2p+1)}$.  

In the insets of Fig.~\ref{fig:lattg2}(b) we plot the correlation 
functions $g^{(2)}(x)$ with $x$ rescaled by the 
corresponding (density-dependent) collective blockade distance $\dcb$. 
For vdW interaction, all the curves for different atomic densities 
then exhibit the same oscillation period $\la \simeq 1.75 \dcb$.  
 
\begin{figure}[t]
\includegraphics[width=7cm]{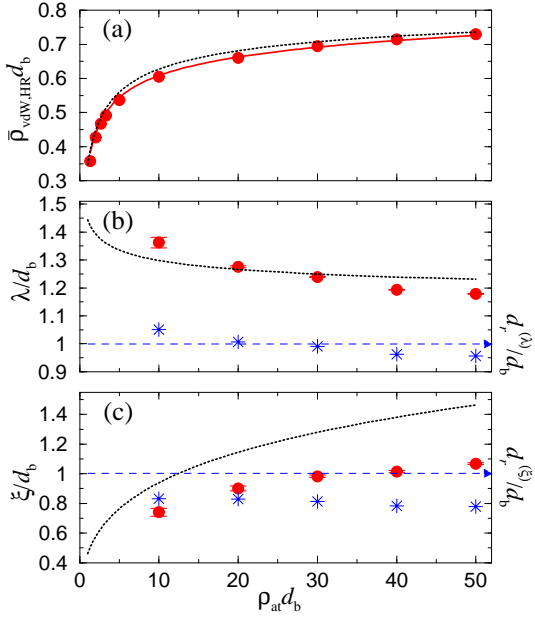}
\caption{%(Color online)
(a)~Average density $\bar{\rho}_{\mathrm{vdW}} = \rho_{\mathrm{at}} \sigav_{rr}$ 
of Rydberg excitations of vdW interacting atoms vs. the atom density 
$\rho_{\mathrm{at}}$ obtained from MC simulations (circles) and 
Eq.~(\ref{sigRRav}) (solid red line). 
Also shown the average excitation density $\bar{\rho}_{\mathrm{HR}}$ of 
Eq.~(\ref{rhoHRav}) for the HR potential of range $\dr=\db$ (dotted black line).
(b)~Oscillation period $\la$ and (c) decay length $\xi$ of spatial 
correlations of excitations $g^{(2)}(x)$ vs. the atom density $\rho_{\mathrm{at}}$.
For vdW interaction potential, the data points (red circles with uncertainty) 
are extracted from the MC simulations; for HR potential, $\la$ and $\xi$ are 
obtained from the solution of Eq.~(\ref{transeqb}). 
The effective range $\dr^{(\la,\xi)}$ (blue stars: right vertical axes) 
of the HR potential is obtained by equating $\la$ (b) and $\xi$ (c) 
for the vdW and HR interactions.}
\label{fig:rholaxi}
\end{figure}

The average (background) density of Rydberg excitations 
$\bar{\rho}_{\mathrm{vdW}} = \rho_{\mathrm{at}} \sigav_{rr}$ 
can also be deduced from the collective excitation picture.
For the average probability of the Rydberg state of vdW interacting 
atoms we obtain 
\begin{equation}   
\sigav_{rr} \approx  \frac{\expv{\Sig_{RR}}}{N_{\mathrm{sa}}} 
= \frac{\expv{\sig_{rr}}}
{ (N_{\mathrm{sa}} - 1) \expv{\sig_{rr}} + 1} \label{sigRRav} , 
\end{equation} 
where $N_{\mathrm{sa}} \approx 1.83 (\rho_{\mathrm{at}} \db)^{12/13}
\sim 2 N_{\mathrm{cb}}$ is the effective number of atoms per superatom
[see Fig.~\ref{fig:rholaxi}(a)]. Note that superatoms overlap.  

%%%%%%%%%%%%%%%%FIGURE%%%%%%%%%%%%%%%%
%%%%%%%%%%%%%%%%%%%%%%%%%%%%%%%%%%%%%%%

\section{Hard-rod potential model}
\label{sec:hr}

We have seen above that the vdW interaction between the atoms in 
Rydberg states results in almost complete blockade of simultaneous 
excitation of two or more atoms within a distance close to $\db$. 
This suggests an analogy between the vdW interaction potential and 
a HR potential of a range $\dr \sim \db$ \cite{Ates2012}. 
The model of incoherently driven atoms interacting via the HR potential 
admits an analytic solution presented below. Our aim is quantitative 
comparison of its predictions with the result of numerical simulations 
for vdW interacting atoms at high densities. 

\subsection{Rate equations model}

In \cite{Hoening2012} we have introduced a rate equations treatment
of the 1D system with complete blockade of Rydberg excitations of 
neighboring atoms. We now extend this formalism to the longer-range
interatomic interactions. To this end, we consider $N$ atoms on a lattice,
with one atom per site, and assume that the atoms excited to the Rydberg 
state $\ket{r}$ interact via the HR potential of range $\dr$: 
$\Delta(\bx_i -\bx_j) = C_{\infty} \Theta(|\bx_i -\bx_j| - \dr)$
with $C_{\infty} \gg w$ and $\Theta(x)$ being the step function.
An atom in state $\ket{r}$ blocks the excitation of 
$N_{\mathrm{r}} = \rho_{\mathrm{at}} \dr$ neighboring atoms on both sides 
along the chain. For each atom, we then have two incoherent processes: 
a pump from $\ket{g}$ to $\ket{r}$ with rate $P$, conditioned upon 
the absence of Rydberg excitations within the range $\dr$, 
$\hat{L}_{\mathrm{p}}^j = \sqrt{P} \, \sig^j_{rg} \prod_{|x_i-x_j| \leq \dr}
\, ( \sig^{i}_{rr} - \mathds{1} )$; 
and a deexcitation from $\ket{r}$ to $\ket{g}$ with rate $D$, 
$\hat{L}_{\mathrm{d}}^j = \sqrt{D} \, \sig^j_{gr}$.
The ratio of the two rates 
$\ka \equiv \frac{P}{D} = \frac{\expv{\sig_{rr}} } {1 - \expv{\sig_{rr}}}$
is obtained from Eqs.~(\ref{sigrr}) with $\de = 0$ as
\begin{equation}
\ka = \frac{|\Om|^2}{|\Om|^2 + \hlf \Ga_r \gamma_{rg}} . 
\end{equation}
The density operator of the system obeys the equation of motion
\begin{equation}
\partial_t \hrho = \sum_j 
\big(2 \hat{L}_{\mathrm{p}}^j \hrho \hat{L}_{\mathrm{p}}^{j\dagger}
- \{\hat{L}_{\mathrm{p}}^{j\dagger}\hat{L}_{\mathrm{p}}^j,\hrho \} 
+ 2 \hat{L}_{\mathrm{d}}^j \hrho \hat{L}_{\mathrm{d}}^{j\dagger}
- \{\hat{L}_{\mathrm{d}}^{j\dagger}\hat{L}_{\mathrm{d}}^j,\hrho \} \big) . 
\label{rhoRE}
\end{equation}
After sufficient relaxation time, the density matrix attains a classical 
form $\hrho = \sum_{\{\mu_j\}} p(\{\mu_j\})\, \ket{\{\mu_j\}} \bra{\{\mu_j\}}$,
where $p(\{\mu_j\})$ is the probability of the $N$-atom configuration 
${\{\mu_j\} \in (g,r)^N}$. In the steady state of the system, 
we have the detailed balance relations
\begin{equation}
\frac{p(\{\mu_j\})}{p(\{\mu_j' \})}=\ka^{ M - M' } ,
\label{eq:detailed-balance}
\end{equation}
where $M \equiv \bra{\{\mu_j\}} \sum_j \sig^{i}_{rr} \ket{\{\mu_j\}}$ 
is the total number of excitations in $\{\mu_j\}$. 
States with the same number of excitations have equal weight 
and the partition function is given by 
$Z_N = \sum_{M=0}^{N} \La(M,N) \ka^{M}$, where 
$\La(M,N) = \binom{N-N_{\mathrm{r}} (M-1)}{M} $ is the number 
of possible arrangements of $M$ excitations on a lattice of $N$ sites,
with any two excitations separated by at least $N_{\mathrm{r}}$ sites. 
The mean number of excitation is given by
$\bar{M} = \frac{1}{Z_N}\sum_{M=0}^N \La(M,N) \ka^{M}$ while 
the average (background) density is $\bar{\rho}_{\mathrm{HR}} = \bar{M}/N$.
In a large system $N \gg N_{\mathrm{r}}$, $\La(M,N) \ka^{M}$ is a 
highly peaked function of $M$; finding its maximum at 
$M_{\mathrm{max}} \simeq \bar{M}$ we obtain the density 
\begin{equation}
\bar\rho_{\mathrm{HR}} = \frac{1}{\dr}\frac{W(\beta)}{1+ W(\beta)} ,
\label{rhoHRav}
\end{equation}
where $\beta = \ka N_{\mathrm{r}}$ and $W(\beta)$ is the Lambert function 
defined via $We^W=\beta$. For $\beta \gg 1$, the density approaches 
$\bar\rho_{\mathrm{HR}} \to 1/\dr$. 
We note that Eq.~(\ref{rhoHRav}) is also obtained in the
classical model of hard rods \cite{hardrods} where $\dr$ is 
the rod length and $\beta$ is a free parameter. Our derivation,
however, yields
\begin{equation}
\beta = \ka \rho_{\mathrm{at}} \dr ,
\end{equation}
which is uniquely determined through the system parameters. 

Next to the boundary at $x=0$, the density of excitations reads
\begin{equation}
\rho_{\mathrm{HR}}(x) = \frac{W(\beta)e^{-W(\beta) x /\dr}}{\dr}
\sum_{k=0}^{\infty} \Theta(x/\dr -k)
\frac{\beta ^k (x/\dr -k)^k }{k!}  , 
\end{equation}
and the spatial correlations are given by 
$g^{(2)}(x)=\rho_{\mathrm{HR}}(x-\dr)/\bar{\rho}_{\mathrm{HR}}$. 
At distances $x > \dr$, the correlations have the form of decaying
spatial oscillations
\begin{equation}
g^{(2)}(x) \simeq 1 + A \cos(2 \pi x/\la + \phi) e^{-x/\xi} , 
\label{g2HR}
\end{equation}
where the oscillation period and decay length, 
\begin{subequations}
\begin{eqnarray}
\la &=& \dr \frac{2\pi}{b} , \\
\xi &=& \dr \ln^{-1} \left(\frac{-b}{W(\beta)\sin(b)}\right) ,
\end{eqnarray}
\end{subequations}
are determined by the solution of the transcendental equation for $b$,
\begin{equation}
\ln \big( W(\beta) \big) + W(\beta) - \ln \left(\frac{-b}{\sin(b)} \right)
 + \frac{b}{\tan(b)} = 0 . \label{transeqb}
\end{equation}

\subsection{Comparison of the vdW and HR potentials}

To quantify the suitability of the HR model \cite{Ates2012} for the
description of the system under the vdW potential, we now compare 
the mean densities of excitations $\bar\rho_{\mathrm{vdW,HR}}$, 
as well as the oscillation periods $\la$ and decay lengths $\xi$ 
of spatial correlations $g^{(2)}(x)$. 

Let us first equate the range of the HD potential to the blockade distance 
of the vdW potential, $\dr = \db$. The average densities of excitations are 
then remarkably close, $\bar{\rho}_{\mathrm{vdW}} \simeq \bar{\rho}_{\mathrm{HR}}$, 
especially at high atomic densities $\rho_{\mathrm{at}} d_{\mathrm{b}} \gg 1$, 
see Fig.~\ref{fig:rholaxi}(a). With increasing the atomic density, 
however, the oscillation period $\la$ of the density wave decreases 
faster for the vdW potential, Fig.~\ref{fig:rholaxi}(b); 
apparently the superatoms having soft boundaries can pack closer 
to each other. The same softness of the vdW potential, compared 
to the HR potential, can explain shorter correlation lengths $\xi$, 
see Fig.~\ref{fig:rholaxi}(c). 

Next, we may assume equal oscillation periods $\lambda$ for both potentials 
and then deduce the corresponding range $\dr^{(\la)}$ of the HR potential, 
which turns out to be close to $\db$ but slowly decreasing with increasing
the atomic density $\rho_{\mathrm{at}}$, Fig.~\ref{fig:rholaxi}(b). 
Alternatively, we equate the correlation lengths $\xi$ and find 
somewhat smaller corresponding range of the HR potential, 
$\dr^{(\xi)} \simeq 0.8 \db$, again slowly decreasing with increasing
the atomic density $\rho_{\mathrm{at}}$, Fig.~\ref{fig:rholaxi}(c).

%%%%%%%%%%%%%%%%FIGURE%%%%%%%%%%%%%%%%
\begin{figure}[t]
\includegraphics[width=8.5cm]{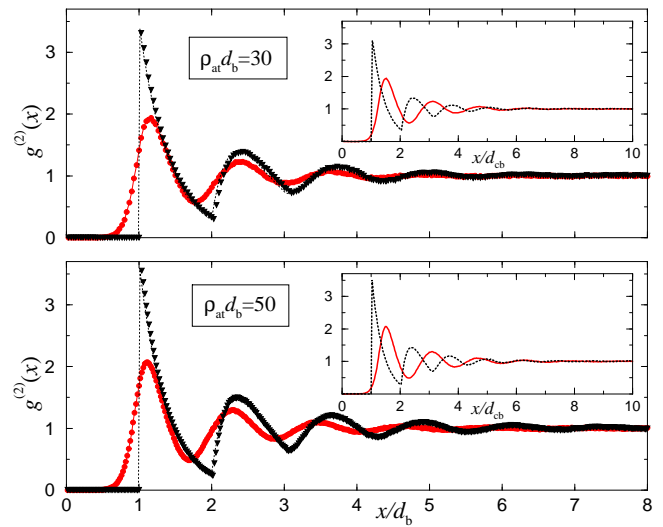}
\caption{%(Color online)
Correlations $g^{(2)}(x)$ of atomic excitations for vdW (circles)
and HR (triangles) potentials obtained from MC simulations
with atom densities $\rho_{\mathrm{at}} \db = 30$ (top) and 
$\rho_{\mathrm{at}} \db = 50$ (bottom). In the insets, $g^{(2)}(x)$ are 
plotted versus $x$ in units of the collective blockade distance $\dcb$.
Atomic parameters are as in Fig.~\ref{fig:lattg2}.}
\label{fig:g2vdWHR}
\end{figure}
%%%%%%%%%%%%%%%%%%%%%%%%%%%%%%%%%%%%%%%

Finally, we have performed MC simulations for $N \gg 1$ strongly-driven 
atoms interacting via the HR potential. In Fig.~\ref{fig:g2vdWHR} 
we compare the the correlation functions $g^{(2)}(x)$ obtained from
our simulations and the corresponding analytic solutions with the 
results for the vdW interacting atoms under the otherwise identical
conditions. We observe qualitatively similar behavior when 
the range $\dr$ of the HR potential is equal to the blockade 
distance $\db$ of the vdW potential. Alternatively \cite{Ates2012}, 
we may associate the range of the HR potential $\dr$ with 
the collective blockade distance $\dcb$ in which case 
$\beta = \ka N_{\mathrm{cb}} = \ka (\rho_{\mathrm{at}}\db)^{12/13}$.
As seen in the insets of Figs.~\ref{fig:g2vdWHR}, the correlation 
functions for vdW and HR potentials are now considerably different,
the corresponding oscillation periods being an approximately 
constant $\la \simeq 1.75 \dcb$ for the vdW potential, and 
a slowly varying $\la \simeq 1.3-1.25 \dcb$ for the HR potential.

\section{Concluding remarks}
\label{sec:conclud}

The system studied in this paper corresponds to an ensemble of 
cold alkali atoms excited to the strongly-interacting Rydberg states 
$\ket{r}$ with principal quantum number $n \sim 50-100$. The resonant
atomic excitation is effected by either direct one-photon (UV) transition
$\ket{g} \to \ket{r}$ or two-photon transition via non-resonant 
intermediate state \cite{Loew2012}. Typical values for the driving field 
Rabi frequency $\Om \sim 10^{5}\:$Hz, together with the relaxation rates 
$\Ga_r \lesssim 0.1\Om$ and $\ga_{rg} \simeq 2\Om$, lead to the blockade 
distance in the range of $\db \sim 5-10\:\mu$m. The low-density 
ensemble of equidistant atoms is trapped in a 1D optical lattice.
At higher densities, regular arrangement of atoms in a lattice 
would play a minor role and our results should hold also in the continuous 
gas of atoms confined in an elongated trap with transverse dimension 
much smaller than the blockade distance. 

To summarize, we have shown that the steady-state probabilities 
of Rydberg excitations in resonantly-driven atomic ensembles 
can exhibit damped spatial oscillations and quasi-crystallization 
due to the strong interactions between the atomic Rydberg states. 
The inverse wavelength and correlation length of these oscillations 
grow with the probability of collective Rydberg excitations of 
superatoms which increases with the atomic density and thereby 
the cooperativity of the excitations. An analytic model based 
on a ``hard-rod'' interatomic potential can serve as a guide to 
understanding the properties of the system under the van der Waals 
potential. 

After sudden switching-off of the driving field, the Rydberg 
quasi-crystal can survive for tens or hundreds of microseconds, 
it can be detected {\it in situ} by spatially-resolved Rydberg 
state ionization \cite{Schwarzkopf2011} or high-resolution 
fluorescence imaging \cite{Schauss2012}.

\begin{acknowledgments} 
Financial support of the DFG through SFB TR49 is acknowledged. 
D.P. is grateful to the University of Kaiserslautern 
for hospitality and support.
\end{acknowledgments}


\begin{thebibliography}{99}
\bibitem{ddpot}
We assume the static DD interaction between the atoms in Rydberg states 
possessing permanent dipole moments induced by an external dc electric 
field \cite{RydAtoms}.
\end{thebibliography}
\end{document}